\documentstyle[12pt]{article}
\def\norm{{v^2 \over \Lambda^2}}
\def\sp#1{{\rm Tr}\biggl( #1 \biggr)}

\def\ra{\rightarrow}
\def\prd#1#2#3{{\it Phys. Rev.} {\bf D#1} #2 (19#3)}
\def\pl#1#2#3{{\it Phys. Lett.} {\bf #1B} #2 (19#3)}
\def\np#1#2#3{{\it Nucl. Phys.} {\bf B#1} #2 (19#3)}

\def\beq{\begin{equation}}
\def\eeq{\end{equation}}

\def\beqn{\begin{eqnarray}}
\def\eeqn{\end{eqnarray}}
\relax
\begin{document}
\begin{titlepage}
\def\ba{\begin{array}}
\def\ea{\end{array}}
\def\thefootnote{\fnsymbol{footnote}}
\begin{flushright}
          UCD-94-6\\
        NUHEP-TH-94-2 \\
	March 1994
\end{flushright}
\vfill
\begin{center}
{\large \bf SEARCH FOR ANOMALOUS COUPLINGS AT \\
$e \gamma $ COLLIDERS\\}
\vfill
	{\bf Kingman Cheung,$^{(a)}$}
	{\bf S.~Dawson,$^{(b)}$}
	{\bf T.~Han,$^{(c)}$}
        {\bf  and G.~Valencia$^{(d)}$}\\
{\it  $^{(a)}$ Department of Physics,
Northwestern University, Evanston, IL~~60208}\\
{\it  $^{(b)}$ Physics Department,
               Brookhaven National Laboratory,  Upton, NY 11973}\\
{\it  $^{(c)}$ Department of Physics,
University of California at Davis, Davis~~CA 95616}\\
{\it  $^{(d)}$ Department of Physics,
               Iowa State University,
               Ames IA 50011}\\
\vfill
     %	{\large \bf ABSTRACT}
\end{center}
\begin{abstract}

We consider the possibility of observing
deviations from the Standard Model gauge-boson  self-couplings
at future $e\gamma$ colliders.
We concentrate on the process $e^-\gamma \ra \nu W^- Z$,
which is particularly sensitive
to the parity violating coupling $g_5^Z$ at high energies.
We find that a 2~TeV $e^+ e^-$ collider operating in the $e^- \gamma$
mode could reach a sensitivity of order $5\times 10^{-3}$ to $g_5^Z$.

\end{abstract}

\end{titlepage}

\clearpage

\section{Introduction}

The Standard Model of electroweak interactions (SM) has
been tested thoroughly in many  experiments. The
only missing ingredients are the top-quark and the Higgs-boson.
Whereas we expect that the top-quark will be found in the
near future, the same cannot be said for the Higgs-boson.
The Higgs-boson in the Standard Model is responsible for
the breaking of electroweak symmetry, and experiments
conducted thus far have not tested directly the energy
scales at which the symmetry breaking is thought to occur.

There are many different physics possibilities that could be
responsible for the breaking of the  electroweak symmetry.
On general grounds we expect that there will be new particles
associated with the symmetry breaking sector, like the Higgs
boson or whatever takes its place. The purpose of new high
energy colliders is to explore the energy scales associated
with the breaking of electroweak symmetry. However, it is
possible that any given machine will be operating at energies
below that which is necessary to excite the lightest new particle.
In that case, we would like to study gauge boson interactions
and try to infer from them what the new physics could be.
This scenario makes it interesting to parameterize the symmetry breaking
sector of the theory in a model independent way by means
of an effective Lagrangian, and to explore
the sensitivity of present and future experiments to the couplings
introduced in this parameterization.

In this paper we will look at the capability of future
$e \gamma$ colliders to study the ``anomalous'' gauge boson couplings.
Such colliders with high energy and high luminosity can potentially
be made from $e^+ e^-$ linear colliders
using the laser back-scattering method  \cite{laser}.
The possibility of observing the SM Higgs boson
at such colliders through the process
$e^- \gamma \rightarrow \nu W^- H$ has been previously discussed,
and it was found that
a 1~TeV $e^+ e^-$ collider operating in
the $e \gamma$
mode should be able to discover a Higgs boson in the intermediate
mass range $60-150$~GeV \cite{cheung1,cheung2}.
Here we assume that there is no Higgs boson (at least not with a mass
that makes it directly accessible to the machine) and investigate the
sensitivity of an $e \gamma$ collider to the three and four gauge-boson
self-interactions \cite{weg}.
We concentrate on the  process $e^- \gamma \ra \nu  W^- Z$
for several reasons. This process depends on only a few
of the anomalous couplings, and at very high energies it singles out
one coupling. The process is thus ideally suited to isolate the
effects of this one coupling, the parity violating but $CP$ conserving
$g_5^Z$. This process also has a relatively small SM background as
we will discuss in section 3.

In Section 2, we summarize the effective Lagrangian formalism that
we use to describe the ``anomalous'' couplings. We derive all the
Feynman rules needed to compute the amplitude for $e^- \gamma \ra \nu W^- Z$.
In section 3 we present our numerical results and discuss
the sensitivity of an $ e^- \gamma $ collider to the anomalous couplings.
Finally, in section 4 we present our conclusions.

\section{Anomalous Couplings}

We start from the minimal effective Lagrangian that describes the
interactions of gauge bosons of an $SU(2)_L \times U(1)_Y$ gauge
symmetry spontaneously broken to $U(1)_Q$.
In this case the gauge boson mass and kinetic energy terms are
\cite{longo}:
\beq
{\cal L}^{(2)}={v^2 \over 4}\sp{D^\mu \Sigma^\dagger D_\mu \Sigma}
-{1\over 2} {\rm Tr}\biggl(W^{\mu\nu}W_{\mu\nu}\biggr)
-{1\over 2}{\rm Tr}\biggl(B^{\mu\nu}B_{\mu\nu}\biggr)\quad ,
\label{lagt}
\eeq
where $W_{\mu\nu}$ and $B_{\mu\nu}$ are
the $SU(2)$ and $U(1)$  field strength tensors given in terms of
$W_\mu \equiv W^i_\mu \tau_i$, by:\footnote{
The Pauli matrices are normalized such that $Tr(\tau_i\tau_j)
=2\delta_{ij}$.}
\beqn
W_{\mu\nu}&=&{1 \over 2}\biggl(\partial_\mu W_\nu -
\partial_\nu W_\mu + {i \over 2}g[W_\mu, W_\nu]\biggr)
\nonumber \\
B_{\mu\nu}&=&{1\over 2}\biggl(\partial_\mu B_\nu-\partial_\nu B_\mu\biggr)
\tau_3.
\eeqn

The matrix $\Sigma \equiv \exp(i\vec{\omega}\cdot \vec{\tau} /v)$, contains the
would-be Goldstone bosons $\omega_i$ that give the $W$ and $Z$ their
mass via the Higgs mechanism, and the $SU(2)_L \times U(1)_Y$
covariant derivative is given by:
\beq
D_\mu \Sigma = \partial_\mu \Sigma +{i \over 2}g W_\mu^i \tau^i\Sigma
-{i \over 2}g^\prime B_\mu \Sigma \tau_3.
\label{covd}
\eeq
The first term in Eq.~\ref{lagt} is the $SU(2)_L \times U(1)_Y$ gauge
invariant mass term for the $W$ and $Z$. The physical masses
are obtained with $v \approx 246$~GeV. This non-linear realization
of the symmetry breaking sector contains the same low energy physics as
the minimal Standard Model when the Higgs-boson is taken to be very
heavy \cite{longo}. It is a non-renormalizable theory
that is interpreted as an effective field theory, valid below
some scale $\Lambda \leq 3$~TeV. The lowest order interactions between
the gauge bosons and fermions are the same as those in the minimal
Standard Model.

In writing the Lagrangian of Eq.~\ref{lagt} we assumed a custodial
$SU(2)_C$ symmetry broken only by the hypercharge coupling $g^\prime$.
Without this assumption, there can be one more term in the lowest
order effective Lagrangian:
\beq
{\cal L}^{(2)\prime}={1 \over 8} \Delta \rho v^2 \biggl[
\sp{\tau_3 \Sigma^\dagger D_\mu \Sigma}\biggr]^2.
\eeq
This term describes deviations of the $\rho$ parameter from
one.  Since, experimentally, $\Delta\rho$ is very small:
$\Delta \rho =(1.5 \pm 2.6)\times 10^{-3} $ \cite{rhop},
 we will  neglect ${\cal L}^{(2)\prime}$.

The ``anomalous'' gauge boson couplings that we want
to study correspond to other $SU(2)_L\times U(1)_Y$ gauge invariant
operators one can write besides those in Eq.~\ref{lagt}. For ``low
energy'' processes, those that occur at energies below the scale
of symmetry breaking $\Lambda$, it is possible to organize the effective
Lagrangian in a way corresponding to an expansion of
high energy scattering
amplitudes in powers of $E^2/\Lambda^2$. The next to leading order
effective Lagrangian that arises in this context
has been discussed at
length in the literature \cite{longo,holdom,bdv,fls,appel}.

We will consider the next-to-leading order effective Lagrangian
containing all terms which conserve the custodial $SU(2)_C$ (up to
hypercharge couplings). In addition, we will
consider a special operator which apart from breaking the
custodial symmetry, violates parity and charge conjugation while
conserving $CP$ \cite{dv}. The process that we discuss in this paper,
$e^- \gamma \ra \nu W^- Z$, is sensitive  to the $\gamma ZWW$
interaction. At very high energies, the amplitude for the process
$\gamma W^- \ra Z W^-$ is enhanced when the $W^-$ and $Z$ are
longitudinally polarized. Roughly, the amplitude gains an enhancement
factor $\sqrt{s}/M_W$ for each longitudinal polarization, and the
largest enhancement is thus obtained when all three gauge bosons
are longitudinally polarized. We could compute this enhanced amplitude
by using the equivalence theorem and replacing each
longitudinally polarized
gauge boson with its corresponding would-be Goldstone boson. For the
largest enhancement we would thus be calculating the amplitude for
$\gamma \omega^- \ra \omega^0 \omega^-$
($\omega^{\pm}, \omega^0$ are the would-be Goldstone bosons
corresponding to the longitudinal components of the $W$ and $Z$).
However, this process will
not take place unless parity is violated (since we are not considering
the possibility of having Wess-Zumino-Witten terms for the effective
theory). This leads us to believe that the process $e^- \gamma \ra \nu W^- Z$
will single out the parity violating couplings at high energy.
For this reason,
out of the many terms that break the custodial symmetry in the
next to leading order Lagrangian, we only consider the one that also violates
parity. This term will yield enhanced interactions at high energies.
There are other terms that violate parity, but they also violate $CP$.
We know that the weak interactions violate parity maximally, but that
observed $CP$ violation is very small. With this in mind, we do not expect
the $P$ violating but $CP$ conserving operator to have any special
suppression factors in general. At the same time, we might expect
the $CP$ violating operators to have smaller couplings than one
would estimate from simple power counting rules. We will not consider
the $CP$ violating operators in this paper.

We are thus left with the following next-to-leading order
effective Lagrangian:
\beqn
{\cal L}^{(4)}\ &=&\ {v^2 \over \Lambda^2}  \biggl\{ L_1 \, \biggl[
\sp{D^\mu\Sigma^\dagger D_\mu \Sigma} \biggr]^2
\ +\  L_2 \,
 \sp{D_\mu\Sigma^\dagger D_\nu \Sigma}
\sp{D^\mu\Sigma^\dagger D^\nu \Sigma}   \nonumber \\
& -& i g L_{9L} \,\sp{W^{\mu \nu} D_\mu
\Sigma D_\nu \Sigma^\dagger}
\ -\ i g^{\prime} L_{9R} \,\sp{B^{\mu \nu}
D_\mu \Sigma^\dagger D_\nu\Sigma} \nonumber \\
& +& g g^{\prime} L_{10}\, \sp{\Sigma
B^{\mu \nu}
\Sigma^\dagger W_{\mu \nu}}\ +\
g {\hat \alpha}
\epsilon^{\alpha \beta
\mu \nu}\sp{\tau_3 \Sigma^\dagger D_\mu \Sigma}
\sp{W_{\alpha \beta} D_\nu \Sigma \Sigma^\dagger}
\biggr\}.
\label{lfour}
\eeqn
The terms with the $L_i$ are the five terms that conserve the custodial
$SU(2)_C$ symmetry.
We explicitly introduce the factor $v^2/\Lambda^2$ in our definition
of ${\cal L}^{(4)}$ so that the
$L_i$ are naturally of ${\cal O}(1)$. The term with $\hat{\alpha}$ is the
one that breaks custodial symmetry and violates parity while conserving
$CP$. With the normalization we have given ${\cal L}^{(4)}$, we expect
$\hat{\alpha}$ to be of ${\cal O}(1)$ in theories without a custodial
symmetry and of order $\Delta\rho$ in theories that have a custodial
symmetry \cite{dv}. For our discussion we will assume that the new
physics is such that the tree-level coefficients of ${\cal L}^{(4)}$ are
larger than the (formally of the same order) effects induced by
${\cal L}^{(2)}$ at one-loop. More precisely, that after using dimensional
regularization and a renormalization scheme similar to the one we used
in Ref.~\cite{bdv}, the $L_i(\mu)$ evaluated at a typical scale around
$\Lambda$ are equal to the tree-level coefficients and that the running
is unimportant for the energies of interest. The physical motivation
for this assumption is that, even if we do not see any new resonances
directly, the effects of the new physics from high mass scales
must clearly stand out if there
is to be any hope of observing them.
When the indirect effects of the
new physics enter at the level of SM radiative corrections, very precise
experiments (as the ones being performed at LEP I) are needed to unravel them.
We are assuming that there will not be any such precision measurements
in the next generation of high energy colliders.

We can find the Feynman rules by going to unitary gauge,
$\Sigma=1$, and expanding the Lagrangians of
Eqs.~\ref{lagt}  and ~\ref{lfour}.
The operator in front of $L_{10}$,
however, contributes terms bilinear in the gauge
boson fields to the effective Lagrangian. It is therefore necessary
to perform additional transformations on the fields $A$ and $Z$ that appear
in our effective Lagrangian to obtain the physical fields.
We follow Holdom \cite{holdom}, and use
the~ ``$*$'' renormalization scheme of
Kennedy and Lynn \cite{kl} which  is defined by the
relation:
\beq
s_Z^2 c_Z^2 \equiv{\pi \alpha^*\over \sqrt{2} G_F M_Z^{2~{\rm phys}}} ,
\quad
\eeq
where $s_Z^{}$ ($c_Z^{}$) is the sine (cosine) of the weak mixing
angle renormalized at the scale of $M_Z^{}$, $s_Z^{}=\sin\theta_W^{}$.
We thus use as input parameters: $G_F$ as measured in muon decay;
the physical $Z$ mass,
$M_Z^{\rm phys}=91.17$~GeV, and $\alpha^*=1/128.8$. In terms of these
quantities we thus use $v =1/\sqrt{\sqrt{2}G_F}$.

Our next-to-leading order amplitudes will get direct contributions
from the terms in ${\cal L}^{(4)}$, as well as contributions from
the renormalization of ${\cal L}^{(2)}$
(including the couplings to fermions that we have not written).
To the order we are working,
this renormalization is accomplished by replacing
the unrenormalized quantities
$e$, $c_\theta$, $s_\theta$ and $M_Z^0$ appearing in ${\cal L}^{(2)}$
in the following way \cite{holdom}:
\beqn
c_\theta&=& c_Z\biggl(1-{2 s_Z^2 e^{*2}
\over s_Z^2-c_Z^2} L_{10}\norm \biggr)  \nonumber \\
s_\theta&=& s_Z\biggl(1+{2 c_Z^2 e^{*2}
\over s_Z^2-c_Z^2} L_{10}\norm \biggr)  \nonumber \\
M_Z^0&=& \biggl(1+e^{*2} L_{10}\norm \biggr) M_Z^{{\rm phys}}\nonumber \\
e&=&\biggl(1-e^{*2} L_{10}\norm \biggr) e^* ~,
\label{renc}
\eeqn
and by re-writing the lowest order Lagrangian in terms of the
physical fields. The only fields that change are those
corresponding to the neutral gauge
bosons:
\beqn
Z& \ra& \biggl(1-e^{*2}L_{10}\norm \biggr) Z \nonumber \\
A& \ra& \biggl(1+e^{*2}L_{10}\norm \biggr) A + {(c_Z^2-s_Z^2)\over s_Z c_Z}
e^{*2}L_{10}\norm Z
\quad .
\eeqn
The relevant Feynman rules are given in the Appendix.

It has become conventional in the literature to parameterize the
three gauge boson vertex $VW^+W^-$ (where $V=Z,\gamma$)
in the following way \cite{hagi}:
\beqn
{\cal L}_{WWV}&= &
-ie_* {c_Z\over s_Z} g_1^Z \biggl(
W_{\mu\nu}^{\dagger} W^{\mu}-W_{\mu\nu} W^{\mu~\dagger}\biggr) Z^\nu
-ie_* g_1^\gamma\biggl(
W_{\mu\nu}^{\dagger} W^{\mu}-W_{\mu\nu} W^{\mu~\dagger}\biggr) A^\nu
\nonumber \\ &&
-ie_* {c_Z\over s_Z} \kappa_Z
W_{\mu}^{\dagger} W_{\nu}Z^{\mu\nu}
-ie_* \kappa_\gamma W_{\mu}^{\dagger} W_{\nu}A^{\mu\nu}
\nonumber \\ & &
-e^*{c_Z \over s_Z} g_5^Z
\epsilon^{\alpha\beta\mu\nu}\biggl(
W_\nu^-\partial_\alpha W_\beta^+-W_\beta^+\partial_\alpha
W_\nu^-\biggr)Z_\mu \quad .
\eeqn
For comparison with the literature we present our results in this form:
\footnote{We agree with the results of
Appelquist and Wu \cite{appel}
when we make the identification, $\alpha_1=\norm L_{10},
\alpha_2=\norm {L_{9R}\over 2},
 \alpha_3=\norm {L_{9L}\over 2}$,
and with those of Holdom \cite{holdom}
when we use
$L_9^{\rm Holdom}=-\norm {(L_{9L}+L_{9R})\over 2}$.
The terms proportional to $L_{9L}$ and $L_{9R}$ agree with those
of Boudjema \cite{boud}.
However, there are several typos in the results of
Falk, Luke, and Simmons \cite{fls}.}
\beqn
g_1^Z&=&1+{e^2\over c_\theta^2}
\biggl({1\over 2 s_\theta^2 } L_{9L}
+{1\over  (c_\theta^2-s_\theta^2)}L_{10}\biggr){v^2\over
\Lambda^2}\nonumber \\
g_1^\gamma&=& 1 \nonumber \\
\kappa_Z&=&1+ e^2\biggl({1\over 2 s_\theta^2
c_\theta^2} \biggl(L_{9L}c_\theta^2
-L_{9R}s_\theta^2\biggr)
+{2 \over  (c_\theta^2-s_\theta^2)}L_{10}
\biggr){v^2\over \Lambda^2}\nonumber \\
\kappa_\gamma&=&1+{e^2 \over s_\theta^2}
\biggl({L_{9L}+L_{9R}\over 2} -L_{10}\biggr)
{v^2\over \Lambda^2}~ \nonumber \\
g_5^Z &=& {e^2 \over s_\theta^2
c_\theta^2}\hat{\alpha}{v^2\over \Lambda^2}.
\label{unot}
\eeqn
The difference between $e$ and $e^*$, $s_\theta$ and $s_Z$ in Eq.~\ref{unot}
is higher order and can be neglected.

The four gauge boson interaction is
derived from
Eqs.~\ref{lagt} and~\ref{lfour} and can be written as:
\beqn
{\cal L}_{WWZA}&=&
-e^{*2} {c_Z \over s_Z} g_1^Z
\biggl( 2 W^+\cdot W^- A\cdot Z
-A\cdot W^+  Z\cdot W^-
-Z\cdot W^+ A\cdot W^-\biggr)\nonumber \\
&& +i\; 2e^{*2} {c_Z \over s_Z} g_5^Z
\epsilon^{\alpha\beta\mu\nu}
W^-_\alpha W^+_\beta Z_\mu A_\nu
\quad
\label{afbg}
\eeqn
It is important to see how, in a gauge invariant description of
anomalous couplings, the coefficients of the three and four gauge
boson couplings are related. Of course, the specific relation
implicit in Eqs.~\ref{unot} and \ref{afbg} may change
if other custodial $SU(2)_C$ violating terms are included in the
Effective Lagrangian.

Our effective Lagrangian formalism breaks down at some scale
$\Lambda \leq 3$~TeV, and this manifests itself in amplitudes that
grow with energy and violate unitarity at some scale related to
$\Lambda$.  In order to stay within the region of validity of our
description, we compute the $J=0$ partial wave for different
amplitudes and require that it be less than one. For the process
$\omega^- \gamma \rightarrow \omega^- \omega^0$,
we find:
\beq
|a_0|_\pm = {1 \over 16 \sqrt{2}} e^* \hat{\alpha}
{\sqrt{s_{\omega\gamma}}\over v}{s_{\omega\gamma}
 \over \Lambda^2}
\eeq
where $\sqrt{s_{\omega\gamma}}$ denotes the center of mass energy of
the $\omega^- \gamma$ system.
For example, if we are interested in studying this process at
$\sqrt{s_{\omega\gamma}}=2$~TeV and we take the scale of new physics to be
$2$~TeV, the requirement that $|a_0|_\pm \leq 1$ leads to the
unitarity bound $|\hat{\alpha}| \leq 9$, or $|g_5^Z| \leq 0.08$.
Notice, however, that this unitarity bound is less strict for
processes at lower energies. This simply corresponds to the
conventional wisdom that the physics associated with electroweak
symmetry breaking must appear at (or below) a scale of a few TeV.

It turns out that the process $\omega^- Z \rightarrow \omega^-
\omega^0$, provides a slightly tighter unitarity
constraint on $\hat{\alpha}$, and it also allows us to compare
it with unitarity bounds on other anomalous couplings.
For example, the lowest order Lagrangian and the term with
$L_1$ in Eq.~\ref{lfour} also contribute to this process. If we again
compute the helicity amplitudes for transverse $Z$ polarizations
and take the $J=0$ partial wave we find:
\beq
|a_0|_{\pm}={1 \over 128 \sqrt{2}}{e^* \over s_Z c_Z}
{\sqrt{s_{\omega Z}} \over v}
\biggl( 1 + 4L_1{s_{\omega Z}^{} \over \Lambda^2}
 + 8 {\hat \alpha} c^2_Z {s_{\omega Z}^{}\over \Lambda^2} \biggr)
\ .
\label{unitra}
\eeq
If we look instead at the amplitude with a longitudinal $Z$ we find:
\beq
|a_0|_L={1 \over 32 \pi}{s_{\omega Z}^{}
 \over v^2}\biggl(1 +{ 16 L_1 \over 3}
{s_{\omega Z}^{} \over \Lambda^2}\biggr)
\ .
\label{unilon}
\eeq
We now adopt the ``naturalness'' argument that each term should be
independently smaller than the unitarity bound. Taking again
$\sqrt{s_{\omega Z}^{}} = \Lambda = 2$~TeV we find from Eq.~\ref{unitra} the
bounds $|\hat{\alpha}| \leq 4.9$ and $|L_1| \leq 7.6$. We see that
unitarity bounds for both couplings are of the same order. However,
Eq.~\ref{unilon} places the much stronger bound
$|L_1| \leq 0.3$.\footnote{This is consistent with the results of a
more general analysis shown in figure 1 of Ref.~\cite{bdv}}
This is well understood from the fact that at high energies one obtains
larger amplitudes for longitudinally polarized gauge bosons. Since
$\hat{\alpha}$ does not contribute to $\omega \omega$ scattering processes,
the constraint on it from unitarity is weaker than that on its counterparts
$L_1$ and $L_2$. On the other hand, $L_9$ and $L_{10}$ will only contribute
to four gauge boson scattering amplitudes when at least two of them are
transversely polarized. This can be seen from the results we presented
in Appendix~A of Ref.~\cite{bdv}. One then finds that the unitarity
bounds on $L_9$ and $L_{10}$ are very loose. For example, from $V V \ra
\omega \omega$ processes with $\sqrt{s_{VV}^{}}=\Lambda=2$~TeV, they are
of the order of $L_{9L,9R,10} \leq 500$.  Bounds on these couplings
from unitarity of $q \overline{q} \ra \omega \omega$ processes
are even weaker.

\section{Phenomenological Studies for $e^- \gamma \ra \nu W^- Z$}

In this section we consider the phenomenology of the process
$e^- \gamma \ra \nu W^- Z$ as it could be studied in future
$e^+ e^-$ or $e^- e^-$ colliders operating in the $e \gamma$ mode.  Within
the SM,
this process has been considered in Ref.~\cite{cheung1}.
The diagrams needed to compute the amplitude
are shown in Figure~1. We evaluate these amplitudes numerically using
helicity amplitude techniques. We have also checked the electromagnetic
gauge invariance  numerically.
This check is done separately for the amplitude corresponding to each
operator since they all must be separately gauge invariant.

We find that
the new interactions proportional to $L_1$ and $L_2$ do not
contribute to this process.
The new interactions proportional to ${\hat {\alpha}},
L_{9L},~L_{9R}$, and $L_{10}$
contribute not only to the  gauge-boson fusion
diagrams, but to all other diagrams as well. This is shown in Figure~1.
The shaded circles on vertices indicate modifications induced by
the new couplings.
This interplay of different couplings induced by the same
gauge invariant operator makes the importance of
our effective Lagrangian  formulation
manifest.\footnote{This process was  recently considered by
Eboli {\it et al.}\cite{eboli}. That paper, however, does not use
a consistent and gauge invariant formalism.}

In Ref.~\cite{dv},
we computed the amplitude for $e^- \gamma\ra \nu  W^- Z $
using the effective $W$ approximation.  The results indicated
that at very high energies,
this process is particularly sensitive to the ${\hat \alpha}$ coupling.
This is, of course, due to the enhanced couplings made possible by
the parity violating nature of this operator. In this section we
compute the amplitude exactly in order to obtain more quantitative
results.

We present all of our numerical results for $e^+ e^-$ colliders.
This involves folding the $e^- \gamma\ra \nu W^- Z $ differential
cross-sections with the energy spectrum of the
back scattered photon. The cross section for $e^+ e^- \ra e^+ \nu W^- Z$
is then,
\beq
\sigma(s_{ee})=\int^{x_{\rm max}}_{x_{\rm min}}
dx F_{\gamma/e}(x){\hat \sigma}_{e \gamma}(xs_{ee}) ,\quad
\eeq
where $s_{ee}$ the squared center of mass energy of the $e^+e^-$ system.
The unpolarized photon luminosity is given by \cite{laser}:
\beq
F_{\gamma/e}(x)={1\over D(\zeta)}\biggl(1-x+{1\over 1-x}
-{4x\over \zeta (1-x)}+{4 x^2 \over \zeta^2(1-x)^2}\biggr)
\eeq
and
\beq
D(\zeta)=\biggl(1-{4\over\zeta}-{8\over \zeta^2}\biggr)
\log(1+\zeta)+{1\over 2} +{8\over \zeta}
-{1\over 2 (1+\zeta)^2}\quad .
\eeq
The parameter $\zeta$ is given by  $\zeta=4 E_0\omega_0/m_e^2$, where
$\omega_0$ is the energy of the incoming laser photon and $E_0$ is the
$e^-$ beam energy.   $x$ is the fraction of the electron
energy carried by the backscattered photon, $x_{\rm max}=\zeta/(1+\zeta)$,
and $x_{\rm min}$ is determined by the kinematical limit $(M_W+M_Z)^2/s_{ee}$.
The value for $\zeta$ is chosen in such a way that the backscattered photon
cannot interact with the incident photon to
produce an unwanted $e^+e^-$ pair.  This requirement
implies $\omega_0 x_{\rm max} \le m_e^2/ E_0$ which gives
$\zeta \le 4.8$. We choose $\zeta=4.8$ to maximize $x_{\rm max}$, and
this yields
$x_{\rm max}=0.83$, $D(\zeta)=1.8$ and $\omega_0=1.25$~eV for a $0.5$~TeV
$e^+ e^-$ collider.

In Fig.~2, we show the total cross section as a function of $\sqrt{s_{ee}}$
for $\Lambda=2$~TeV.
We have adopted the usual ``naturalness'' assumption,
that the process can
independently place bounds on the different anomalous couplings that
contribute to it. This means that we only look at one anomalous coupling
at a time and ignore possible interference between terms with
different  anomalous couplings.  There are four curves in the plot:
the solid one corresponds to the lowest order amplitude only
(that arising from ${\cal L}^{(2)}$);
the dashed curve corresponds to the ${\cal L}^{(2)}$
amplitude plus the one obtained
from ${\cal L}^{(4)}$ with $\hat \alpha=5$;  for the dotted and dash-dotted
curves we use the ${\cal L}^{(2)}$
amplitude plus that obtained from ${\cal L}^{(4)}$
with $L_{9L}=5$  and $L_{9R}=5$ respectively.

Since $L_{10}$ is proportional to the parameter $\epsilon_3$
measured at LEP I, we have a very strong constraint on it. Including
all the
LEP results, as well as all the relevant low energy data, Altarelli
finds $\epsilon_3 = (3.5 \pm 2.8)\times 10^{-3}$ \cite{rhop}. This
corresponds to
\beq
L_{10}(M_Z) = -1.31 \pm 1.05.
\label{altb}
\eeq
Our calculation in this
paper is at tree-level only, so it is not appropriate to
include the running of $L_{10}$. That running, however, would be a
very small effect \cite{bdv}. Since the process $e^- \gamma
\rightarrow \nu  W^- Z$ is only sensitive to much larger values of $L_{10}$
than allowed by Eq.~\ref{altb},
we will not consider this operator in any detail.

We see that, as expected, the cross  section is increasingly more sensitive
to the anomalous couplings at higher energies.
Our result confirms that the amplitude from the term with coupling
${\hat \alpha}$ grows more rapidly with energy than the other terms.
As we argued before, the parity violating nature of this operator allows
it to generate a ``maximally'' enhanced amplitude growing as
$(s_{WZ}^{})^{3/2}$, whereas the amplitudes from the operators with
the $L_{9L}$, $L_{9R}$ couplings only grow as $s_{WZ}^{}$
asymptotically. The relative sensitivity of this process to $\hat{\alpha}$
is thus enhanced at high energies.

In accordance with the ``naturalness'' assumption that we have adopted,
we compute the cross section numerically by looking at one anomalous
coupling at a time, and  further parameterize it as:
\beq
\sigma = \biggl( c_0 + c_1 L_i + c_2 L_i^2 \biggr)~{\rm fb},
\label{csp}
\eeq
where $L_i$ generically represents an anomalous coupling
constant.
The first term, $c_0$, is the contribution from the lowest order
Lagrangian, which is numerically equal to what would be found in the
Standard Model at {\bf tree} level. This is because
the Higgs boson in the Standard Model does not enter
 the calculation of  this process at tree level.
The $c_1$ term corresponds to the interference between
the lowest order amplitude and the amplitude due
to the coupling $L_i$.  The $c_2$ term is the contribution
to the cross section from the square of the amplitude corresponding to the
coupling $L_i$. Notice that since we
only consider one anomalous coupling at a time, there is no interference
between amplitudes corresponding to different anomalous couplings.

We first consider the case of  an $e^+ e^-$  linear collider
with $ \sqrt{s_{ee}}= 0.5$~TeV. This $e^+e^-$ center of mass energy
yields a spectrum for $\sqrt{s_{e\gamma}}$ that peaks at about 0.45~TeV.
The coefficients  for Eq.~\ref{csp} are presented in Table~\ref{t: coel}.
To roughly simulate the detector performance, we have
used the following minimal acceptance cuts,
\beq
|\cos\theta_V |< 0.9, \quad p_T^{}(WZ) > 15~{\rm GeV},
\label{cutsi}
\eeq
where the $\theta_V$ cut is applied to both
the polar angle of $W$ and $Z$, and $ p_T^{}(WZ)$ is
the transverse momentum of the gauge-boson pair. This cut insures
that there is
sufficient missing energy in the event to suppress possible backgrounds.
A more detailed discussion of reducible backgrounds will be deferred to the
end of this section.
For each coefficient, the first row in the table
corresponds to the numbers obtained without cuts
and the second  row gives the results
with the cuts. The constant $c_0$, which gives the minimal SM rate,
is found at $\sqrt{s_{ee}}=0.5$ TeV to be:
\[ c_0 = \left  \{ \begin{array}{ll}
                69.3  & \mbox{no cuts} \\
                42.9  & \mbox{with cuts}.\qquad
                           \end{array}
               \right. \]
\begin{table}[htb]
\centering
\caption[]{Coefficients in Eq.~\ref{csp} for $\sqrt{s_{ee}}=0.5$~TeV.
Cuts are given in Eq.~\ref{cutsi} }
\begin{tabular}{|l|c|c|c|c|c|} \hline
  & $\hat{\alpha}$ & $L_{9L}$ & $L_{9R}$  & $L_{9L}=L_{9R}$  & $L_{10}$  \\
\hline
$c_1$
& $0.324 $ & $0.746 $ & $ 0.281 $ &
      $1.03 $ & $0.312 $ \\ \hline
$c_1$ with  cuts
& $0.209$ & $0.444$ & $0.169$ &
      $0.613 $ & $0.187$ \\ \hline
$c_2$
& $3.80 \times 10^{-2} $ & $4.30 \times 10^{-3}$ & $7.60 \times 10^{-4}$ &
      $7.33 \times 10^{-3}$ & $2.99 \times 10^{-3}$ \\ \hline
$c_2 $ with cuts
& $2.81 \times 10^{-2}$ & $2.58 \times 10^{-3}$ & $5.22 \times 10^{-4}$ &
      $4.63 \times 10^{-3}$ & $1.99 \times 10^{-3}$ \\
\hline
\end{tabular}
\label{t: coel}
\end{table}

Although  the dynamical distributions with anomalous couplings
are not significantly different from  those of the SM at
$\sqrt{s_{ee}}=0.5$ TeV, one can try to
measure deviations from the SM results
based on the total event rate by  computing
the statistical  significance,  which is taken as
\beq
S = {  \rm signal  \,  \, events  \over {\sqrt  {\rm background \,  events} }}.
\eeq
In our case, the ``signal'' events correspond to the second and third
terms in Eq.~\ref{csp}, and the ``background'' events to the minimal
SM rate (the first term in Eq.~\ref{csp}).
A $3 \sigma$ significance resulting from the anomalous couplings
is given in Fig.~3 for  $\hat \alpha, L_{9L}$, $L_{9R}$, and $L_{9L}=L_{9R}$,
respectively,
versus the integrated luminosity at $\sqrt{s_{ee}}=0.5$ TeV.
There is comparable sensitivity
to   $\hat \alpha$ and $L_{9L}$, and
it seems quite feasible for an $e^+e^-$ machine
with $\sqrt{s_{ee}}=0.5$ TeV running in the $e\gamma$ mode, to
reach a sensitivity of order of $10~(\Lambda /2~{\rm TeV})^2$
to these couplings with
an integrated luminosity of $10-20$~fb$^{-1}$.
Such a machine  is  significantly less sensitive  to $L_{9R}$, however.

To demonstrate the significant increase in sensitivity to $\hat \alpha$ at
higher energies, we now  consider a 2 TeV $e^+e^-$ collider.
As we have discussed before, we expect the amplitude proportional to
$\hat \alpha$ to grow faster with energy than the other amplitudes.
This feature is clearly demonstrated in
Fig.~4(a), where the invariant mass distribution of the
gauge boson pairs $M(WZ)$ is shown at $\sqrt{s_{ee}}=2$~TeV
for $\hat \alpha=3~(\Lambda /2~{\rm TeV})^2$ and
$L_{9L}=10~(\Lambda /2~{\rm TeV})^2$.  The
result for $L_{9R}$ is similar to that for
$L_{9L}$ and is not shown here.
We see the increasing importance of the  $\hat \alpha$ contribution as
$M(WZ)$ increases. The effect of the contribution from the $L_9$ term
is mostly to change the overall normalization.
Figure 4(b) shows the  $\cos\theta_W$ distribution at
$\sqrt{s_{ee}}=2$~TeV, where $\theta_W$ is the polar angle of the $W$-boson
with respect to the $e^-$ beam direction.   The SM curve peaks sharply at
$\cos\theta_W=-1$, this is due to the emission of the $W$-boson
from the photon leg.
The $\hat \alpha$ anomalous interaction produces the $W$-boson more
in the central region.  We thus find that the contribution of $L_9$
to the central region is relatively less important than that of $\hat\alpha$.

Since the effect of the $L_9$ interactions is mainly an
overall normalization, to maintain a high sensitivity to
the
$L_9$ coefficients we take moderate acceptance cuts which will
retain most of the signal:
\beq
|\cos\theta_V |< 0.9, \quad p_T^{}(WZ) > 30~{\rm GeV}.
\label{cutsii}
\eeq
Here the $p_T(WZ)$  cut  is optimized to suppress reducible backgrounds
from other sources.
On the other hand,  the results in Fig.~4  imply the possibility of
improving the sensitivity to $\hat \alpha$ by a set of  more stringent cuts:
\beq
|\cos\theta_V |< 0.8, \quad  p^{}_T(WZ) > 30~{\rm GeV},
\quad M(WZ) > 0.5~{\rm TeV}.
\label{cutsiii}
\eeq

The coefficients of Eq.~\ref{csp} for $\sqrt{s_{ee}}=2$~TeV are  given
in Table~\ref{t: coeh}. Again, the first row  corresponds
to the case with no cuts, the second row presents numbers with
the cuts of Eq.~\ref{cutsii}, and the third row shows the numbers with the
cuts of Eq.~\ref{cutsiii}.
The minimal SM rates are
\[ c_0 = \left  \{ \begin{array}{ll}
                936  & \mbox{no cuts} \\
                209  & \mbox{with cuts Eq. \ref{cutsii}} \\
                38.8  & \mbox{with cuts Eq. \ref{cutsiii}} \quad .
                           \end{array}
               \right. \]
\begin{table}[htb]
\centering
\caption[]{Coefficients in Eq.~\ref{csp} for $\sqrt{s_{ee}}=2$~TeV.
Cuts are given in Eqs.~\ref{cutsii} and \ref{cutsiii}}
\begin{tabular}{|l|c|c|c|c|c|} \hline
& $\hat{\alpha}$ & $L_{9L}$ & $L_{9R}$  & $L_{9L}=L_{9R}$  & $L_{10}$  \\
\hline
$c_1$
& $1.03 $ & $10.0 $ & $3.46 $ &
      $13.5 $ & $4.79$ \\ \hline
$c_1$ with cuts Eq. (\ref{cutsii})
&  0.331  &  1.97  &  0.712  &  2.68  & 1.03 \\ \hline
$c_1$ with cuts Eq. (\ref{cutsiii})
& $0.144 $ & $0.393 $ & $0.121$ &
      $0.515$ & $0.225$ \\ \hline
$c_2$
& $9.22 $ & $ 0.165 $ & $0.0425 $ &
      $ 0.294 $ & $ 0.195 $ \\ \hline
$c_2$ with cuts Eq. (\ref{cutsii})
& 6.13  & 0.0565  &  0.028 &  0.137  & 0.122 \\   \hline
$c_2$ with cuts Eq. (\ref{cutsiii})
& $5.12 $ & $0.0282$ & $0.0187$ &
      $0.0820$ & $0.0791$ \\
\hline
\end{tabular}
\label{t: coeh}
\end{table}

At  $\sqrt{s_{ee}}=2$~TeV,  we find that the contribution linear in
${\hat \alpha}$ has a much smaller
coefficient than the quadratic term.
This is consistent with the fact that within
the effective $W$ approximation, this interference term vanishes \cite{dv}.
In the full calculation it is suppressed by $M_W^2/s_{WZ}^{}$ with
respect to the quadratic term.
This effect is not so apparent at the lower energy,
$\sqrt{s_{ee}}=0.5$~TeV.  At $\sqrt{s_{ee}}=2$~TeV,
our results  agree well with those obtained
using the effective $W$ approximation.

A  $3 \sigma$ significance
resulting from
the anomalous couplings
 $\hat \alpha, L_{9L}$, and $L_{9R}$, respectively
at $\sqrt{s_{ee}}=2$~TeV is given in Figs.~5 and 6
for  the cuts of  Eq.~\ref{cutsii} and Eq.~\ref{cutsiii},
as a function of the integrated luminosity.
The curves in (a) correspond to
positive values of the anomalous couplings, while those in (b)
to negative values.
It is seen that the negative values are always more difficult
to probe. In Fig. ~5(b), the solid curve for
$\hat\alpha$ and the dotted curve on the top for $-L_{9L}=-L_{9R}$
are for $+3\sigma$ effects as are those in Fig. 5(a).
There are also three contour-like curves labelled
by $-3\sigma$, which reflect the fact that
$c_1L_i+c_2 L_i^2 < 0$ in Eq. ~\ref{csp}.
This indicates that for certain negative values of the
$L_i$, there are significant
cancellations between the linear and
quadratic terms.  However, with the more
stringent cuts of Eq.~\ref{cutsiii}, we see from Fig.~6(b) that
this effect disappears.  We therefore consider the cancellation to be
an accident.
{}From Fig.~5, we find that the process
$e \gamma \ra \nu W Z$
 could probe $L_9$'s at a level of  $2-5 ~(\Lambda /2~{\rm TeV})^2$
 with an integrated luminosity of
about 100 fb$^{-1}$ and that it
is more sensitive to $\hat\alpha$.
The cuts of Eq.~\ref{cutsiii}  are effective in suppressing the
lowest order contribution (SM),
which is roughly reduced by  a factor of 25
while  the coefficient $c_2$ for the ${\hat \alpha}$ amplitude
is reduced by less than a factor of 2.
We see from Fig. 6 that the coefficient
$\hat \alpha$ can be probed  here to a level less than $1~(\Lambda /2~{\rm
TeV})^2$.

So far we have based our discussion on the excess of
events above the SM prediction. Without distinctive features in some
dynamical distribution,
it might be hard to convincingly establish a signal
for new physics.  Given the parity violating nature of the operator multiplying
$\hat{\alpha}$ we might try to enhance the sensitivity to this
coupling by studying a parity odd observable, such as a correlation
$\vec{p}_e \cdot (\vec{p}_W \times \vec{p}_Z)$. Notice, however, that
there is no reason for this correlation to vanish in the minimal  SM
 where parity is violated maximally. As a measure of
the size of this correlation we can construct an associated
forward-backward asymmetry. We find that for a $500$~GeV collider
the interference term between $\hat{\alpha}$ and the lowest order
amplitude can contain an asymmetry of order $10\%$ for a right-handed
photon. The choice of a right-handed photon is motivated by the
fact that this polarization enhances the relative size of the
$\hat{\alpha}$ term in the total
cross-section in the effective $W$ approximation \cite{dv}.
However, since this interference term is much smaller than the lowest
order term, the measurable asymmetry gets diluted to a negligible level.
Since the relative importance of the interference between the
lowest order amplitude and the
amplitude proportional to $\hat{\alpha}$ decreases
as  the center of mass energy increases, and since a correlation of the type
$\vec{p}_e \cdot (\vec{p}_W \times \vec{p}_Z)$ appears in this
interference, we expect the asymmetry
to be even smaller in a $2$~TeV collider.
Even though  it is very difficulty to single out
the $\hat{\alpha}$ term with a parity
odd observable, we can still study this contribution at high energies.
As we discussed  earlier,  the existence of  ${\hat \alpha}$ significantly
enhances the cross section in the high $M(WZ)$  and  central region,
making the separation of the  ${\hat \alpha}$  contribution  feasible,
as can be seen from Fig.~4.
Due to a Jacobian peak in the $WZ$ two-body  kinematics,
similar enhancements  in  the $p_T(W)$ or  $p_T(Z)$ distributions
also appear in  the high energy region.

The existing data from the CERN $p\bar p$ collider \cite{cern} and
LEP I \cite{lepi}  only constrain the anomalous couplings $L_{9R,9L}$ rather
weakly.
The bound is $|\Delta \kappa| \simeq 1-2$, which translates into
$L_{9R,9L}  <  (300 - 600)~(\Lambda /2~{\rm TeV})^2$.  The Fermilab Tevatron
with an integrated luminosity of  100 fb$^{-1}$ can provide  comparable
results to this bound~\cite{bauretal}.
Ref. \cite{fls} finds that the LHC will be sensitive
to  values $L_{9R}  >  61~(\Lambda /2~{\rm TeV})^2$
 and $L_{9R} < -63~(\Lambda /2~{\rm TeV})^2$;
$L_{9L} > 5~(\Lambda /2~{\rm TeV})^2$
 and $L_{9L}  < -9~(\Lambda /2~{\rm TeV})^2$.
It is well known that LEP II will provide a good environment to
study the three gauge-boson anomalous couplings \cite{lepii} at
a sensitivity of  $|\Delta \kappa| \simeq  4\times 10^{-2}$.
This corresponds to $L_{9R,9L}  \simeq  12~(\Lambda /2~{\rm TeV})^2$.
 There are several
studies in the literature for an $e^+ e^-$ machine
with $\sqrt{s_{ee}}= 0.5$ TeV concentrating
on the $e^+ e^- \ra W^+ W^-$ mode. The claim is that a sensitivity of order
$5\times 10^{-3}$ to $\Delta \kappa_\gamma$ can be reached with
50 fb$^{-1}$ integrated luminosity \cite{miya}.
This would correspond to potential sensitivities of about
$(1-2)~(\Lambda /2~{\rm TeV})^2$ to the couplings $L_{9L}$, $L_{9R}$, and
$L_{10}$
(we have made no attempt to study differences between these three
couplings in that process).  This machine running in the $e\gamma$
mode is also considered in Ref.~\cite{godfrey} as a means
of searching
for anomalous couplings,
and  it is found that  with 50 fb$^{-1}$ a 7\% sensitivity to $\Delta \kappa$,
corresponding to  $L_{9R,9L}  \sim 20$,  is feasible. In contrast,
not many studies
have been done in the literature concerning the coupling $g_5$
(or  ${\hat \alpha}$).
It was found in Ref.~\cite{dv} that the process
$e^+ e^- \ra W^+ W^-$ may be sensitive
to $g_5$ at a high energy
$e^+e^-$ machine  if  highly polarized electron beams can be
obtained.

Finally, let us comment on the event reconstruction and other
possible backgrounds.  With the leptonic decays of $W \ra l\nu, Z \ra l\bar l$,
there is little background to anomalous coupling signals except
for the lowest order $WZ$ production which we have systematically
included in our discussion ($c_0$ in Eq.~\ref{csp}).
In estimating the sensitivities in our figures,
we have implicitly assumed the full use of  the  hadronic decay modes,
in which the four jets in pairs reconstruct to $M_W$ and $M_Z$.
The inclusion of the hadronic modes increases
the event rate significantly and will make  a better $M(WZ)$ mass
reconstruction possible as well. It would  be ideal to have a good
hadronic energy resolution (better than $M_Z-M_W \sim10$ GeV)
in order to be able to distinguish the
$W$ or $Z$ decays via the di-jet mass. If possible, this would
essentially  eliminate other backgrounds.  Given the fact that this
hadronic energy resolution may be difficult to achieve, one has to
make use of other kinematic cuts to reduce potential backgrounds.  The
potential backgrounds are $e^-\gamma \to W^+W^-e^-,\,ZZe^-,\,b \bar t \nu$,
and $t\bar t e^-$, which have all been analysed in Ref.~\cite{cheung2} in the
search for the SM Higgs boson.   Among the backgrounds, the $ZZe^-,\, b\bar
t\nu$, and $t\bar te^-$ are at least an order of magnitude smaller than the
SM process $e^-\gamma\to W^-Z\nu$ and they are sufficiently different
from our signal process in terms of the final state particles and kinematics.
However,  the cross section for the process $e^-\gamma\to W^+W^-e^-$
is about an order of magnitude larger than the SM
$WZ$ process.   Based on the fact that the major contribution to
$e^-\gamma\to W^+W^-e^-$ is from the almost on-shell photon exchange
$\gamma \gamma \to W^+W^-$, we can
substantially reduce it by pushing the photon propagator away from the pole.
Our $p_T(WZ)$ cut is designed for this purpose.  Furthermore, in our $WZ$
signal the transverse momentum of the boson pair is balanced by the
 missing transverse
momentum (essentially from the $\nu$); whereas the $e^-\gamma\to W^+W^-e^-$
background will have a visible $e^-$ with large transverse momentum.  We can
therefore further reduce this background by requiring not only large $p_T(WZ)$
but also, at the same, time vetoing the hard electrons \cite{cheung2}.
Overall, the only  significant irreducible background
for our study of anomalous couplings is the lowest order SM process
$e^-\gamma \to W^-Z\nu$, which we have included.

\section{Conclusions}

If the electroweak symmetry breaking sector is strongly coupled, and
no light resonances are found,
then one expects deviations of the gauge-boson self-interactions from the
SM predictions.
An $e\gamma$ collider operating at $\sqrt{s_{ee}}> 0.5$~TeV can
provide important input into our understanding of the nature
of electroweak symmetry breaking.   Such a collider
would provide more precise measurements  of  $L_{9L,9R}$ than the LHC
via the process  $e^- \gamma \ra \nu W^- Z$.  Due to the gauge
structure and  the relatively large contribution of the gauge-boson
self-interactions to this process,
it is as useful in studying anomalous couplings
as the lower order process $e^- \gamma \ra \nu W^-$\cite{godfrey}.
Although the process  $e^+ e^- \ra W^+ W^-$ is better  to  study
the coefficients  $L_{9L,9R}$,  the process
 $e^- \gamma \ra \nu W^- Z$ is very sensitive to the coupling $\hat \alpha$,
especially at higher energies.
We find that
an $e^+e^-$ machine operating in the $e \gamma$ mode
with $\sqrt{s_{ee}}=0.5$ TeV can place bounds of order
$\hat \alpha~ < ~10 ~(\Lambda /2~{\rm TeV})^2$
with 20 fb$^{-1}$ integrated luminosity.
Similarly,
it is possible at a  2 TeV $e^+e^-$ collider running
in the $e\gamma$ mode to place the bound
$\hat \alpha~<~.6 ~(\Lambda /2~{\rm TeV})^2$,
which corresponds to $g_5^Z < 5 \times 10^{-3}~
(\Lambda/2~{\rm TeV})^2$.
Observation of enhanced production of  longitudinal $WZ$ pairs
 in the high $M(WZ)$ region and at central $\cos\theta_W$
could  be a  direct indication for a new interaction
of the form proportional to $\hat \alpha$.

\section*{Acknowledgments}

The work of S. Dawson is
supported by the U.S. DOE under contract DE-AC02-76CH00016.
K.C. is supported by the U.S. DOE grant  DE-FG02-91-ER40684.
T.H. is supported in part  by  the DOE grant  DE-FG03-91ER40674
and in part  by  a UC-Davis Faculty Research Grant.

\appendix

\section{ Feynman Rules}

The three gauge boson vertex in unitary gauge
 for $Z^\sigma(k)\rightarrow
W^{+~\mu}(p^+)+W^{-~\nu}(p^-)$ can be written
\beqn
iC^{\mu \nu\sigma}_Z(p^+,p^-,k)&\equiv &
i e_* {c_Z\over s_Z} \biggl\{ \biggl[(p^+-p^-)^\sigma g^{\mu\nu}
   +2p^{-\mu} g^{\nu\sigma}-2p^{+\nu} g^{\mu\sigma}\biggr]g_1^Z
\nonumber \\ &&
+\biggl[ k^\mu g^{\nu\sigma}-k^\nu g^{\mu\sigma}\biggr]\kappa_Z
+i  g_5^Z
\epsilon^{\mu \sigma \nu\rho} (p^+-p^-)^{\rho}\biggr\}
\quad .
\eeqn
The momentum is defined
such that $k$ is incoming and $p^+, p^-$ are outgoing.

Similarly the three gauge boson vertex in unitary gauge
 for $\gamma^\sigma(k)\rightarrow
W^{+~\mu}(p^+)+W^{-~\nu}(p^-)$ can be written
\beqn
iC^{\mu \nu\sigma}_\gamma(p^+,p^-,k)&\equiv &
i e_* \biggl\{ \biggl[(p^+-p^-)^\sigma g^{\mu\nu}
   +2p^{-\mu} g^{\nu\sigma}-2p^{+\nu} g^{\mu\sigma}\biggr]g_1^\gamma
\nonumber \\ &&
+\biggl[ k^\mu g^{\nu\sigma}-k^\nu g^{\mu\sigma}\biggr]\kappa_\gamma
\biggr\}
\quad
\eeqn

The four gauge boson vertex in unitary gauge for $W^+_\mu W^-_\nu
\gamma_\kappa Z_\lambda$ is,
\beq
i{\tilde C}^{\mu \nu\kappa\lambda}_{WWZA}=
-ie_*^2 {c_Z\over s_Z}
\biggl\{ \biggl[2 g_{\kappa\lambda} g_{\mu\nu}
-g_{\kappa\mu}g_{\lambda\nu}-g_{\kappa\nu} g_{\lambda
\mu}\biggr]g_1^Z -2 i g_5^Z
\epsilon^{\nu \mu \lambda \kappa} \biggr\}
\quad ,
\eeq
where all particles are incoming.

The renormalization of the couplings given in Eq.~\ref{renc} also affects
the fermion- gauge boson vertices. For $e^+ \nu \ra W^+_\mu$ we find:
\beq
i\Gamma^\mu =
-i{e_*\over 2 \sqrt{2} s_Z }\biggl[1-{e_*^{2}\over s_Z^2-c_Z^2}
L_{10}{v^{2}\over \Lambda^2}\biggr]
\gamma_\mu(1-\gamma_5)
\quad .
\eeq
For a fermion with charge $eQ_f$ and isospin
$T_{3f}=\pm 1$, the ${\overline f} f\ra Z_\mu$ coupling is:
\beq
i\tilde{\Gamma}_\mu =  -i{e_*\over 4 s_Z c_Z}
\gamma_\mu\biggl[
R_e (1+\gamma_5) +L_e (1-\gamma_5)\biggr ]
\eeq
where
\beqn
R_f&=&-2Q_f \biggl[ s_Z^2
- {e_*^2 \over c^2_Z - s^2_Z} L_{10}{v^2\over \Lambda^2}
\biggr]\nonumber \\
L_f&=&R_f+T_{3f}  \quad .
\eeqn
The ${\overline f}f A$ vertices have the same form as they do in
the minimal standard model but they are now given in terms of $e_*$.

\newpage

\noindent{\bf FIGURE CAPTIONS}

\begin{enumerate}

\item  Diagrams contributing to the process $e^- \gamma\ra \nu  W^- Z $.
The dashed circle represents three and four gauge boson couplings arising
from ${\cal L}^{(2)}+{\cal L}^{(4)}$ as given in the appendix. The full
circle represents the renormalized fermion-gauge-boson coupling as given
in the appendix.

\item  Total cross section for $e^+ e^-\ra e^+ \nu W^- Z $
(in the $e^- \gamma$ mode) as
a function of $\sqrt{s_{ee}}$. The solid curve is the result of the lowest
order effective Lagrangian. To this lowest order result, we have
added the contributions from non-zero couplings with
$\Lambda=2$ TeV: The dashed line corresponds to $\hat{\alpha}=5$ and
the dotted and dashed lines correspond to $L_{9L}, L_{9R}=5$,
respectively. The effect  of $L_{10}$ subject to the constraint  of Eq.~
\ref{altb}
is so small that  it cannot be distinguished from the lowest order SM result.

\item $3\sigma$ sensitivity of an $e^+ e^-$ collider at $\sqrt{s_{ee}}=0.5$ TeV
(operating in the $e^- \gamma$ mode) to $\hat{\alpha},  L_{9L}$
and $L_{9R}$ with the cuts Eq.~\ref{cutsi}.
The curves are  shown as a function of integrated luminosity.
We set $\Lambda = 2$~TeV.

\item  Differential cross sections
(a) $d  \sigma /dM(WZ)$ and (b) $d\sigma /d(\cos\theta_W)$
for  $e^+ e^-\ra e^+ \nu W^- Z $.
$\theta_W^{}$ is the polar  angle between the $W^-$
and the $e^-$ in the lab frame and the total center of mass energy
$\sqrt{s_{ee}}= 2$ TeV.
The curves show the result for the lowest order effective
Lagrangian (solid, labeled as SM),  for $\hat{\alpha}=3$
(dashes), and for  $L_{9L}$  (dots)  respectively,
for $\Lambda=2$~TeV.

\item Same as Figure~3 but for $\sqrt{s_{ee}}=2$~TeV with
the set of cuts Eq.~\ref{cutsii}.
(a) for positive values of anomalous couplings and (b) for negative.

\item Same as Figure~3 but for $\sqrt{s_{ee}}=2$~TeV
with the set of cuts Eq.~\ref{cutsiii}.
(a) for positive values of anomalous couplings and (b) for negative.

\end{enumerate}


\begin{thebibliography}{999}
\bibitem{laser} {I.~F.~Ginzburg {\it et. al.}, \np{228}{285}{83};
I.~F.~Ginzburg {\it et. al.}, Nucl. Instrum. \& Methods {\bf 205} 47 (1983);
{\it ibid} {\bf 219} 5 (1984); V.~I.~Telnov,
Nucl. Instrum. \& Methods {\bf A294} 72 (1990);
D.~Bauer, D.~Borden, D.~ Miller, and J.~Spence, {\it 9th International
Workshop on Photon-Photon Collisions}, San Diego, CA.~~1992;
.}

\bibitem{cheung1}{Kingman Cheung, \np{403}{572}{93};
S.~Moretti, DFTT 79/93}.

\bibitem{cheung2}Kingman Cheung, \prd{48} {1035}{93}.

\bibitem{weg} {S.~Godfrey, G.~Couture, and P. Kalyniak,
{\it Beyond the Standard Model}, Ames, Iowa, 1988;
S. Godfrey, G.~Couture, and P. Kalyniak, {\it 11th Annual
Montreal-Rochester-Syracuse-Toronto Meeting}, Syracuse, N.Y., 1989;
E.~Yehudai, \prd{41} {33} {90};
G.~Couture, S.~Godfrey, P.~Kalyniak, \pl{218}{361}{89};
G.~Belanger and F.~Boudjema, \pl{288}{201}{92};
D.~Espriu and M.~Herrero, \np{373}{117}{92};
A.~Dobado and J.~Urdiales, \pl{292}{129}{92};
R.~Casalbuoni {\it et. al.}, \np{409}{257}{93};
D.~Choudhury and F.~Cuypers, MPI-Ph/93-98
.}

\bibitem{longo}{T.~Appelquist and C.~Bernard, \prd{22}{200}{80};
                A.~Longhitano, \np{188}{118}{81}.}

\bibitem{rhop}{G.~Altarelli, Plenary Talk given at the EPS Conference on
High Energy Physics, Marseille, France, July 1993, CERN-TH-7045/93.}

\bibitem{holdom}{B.~Holdom, \pl{258} {156} {91}.}

\bibitem{bdv}{J.~Bagger, S.~Dawson and G.~Valencia, \np{399}{364}{93}.}

\bibitem{fls}{A.~Falk, M.~Luke, and E.~Simmons, \np{365}{523}{91}.}

\bibitem{appel}{T.~Appelquist and G.-H.~Wu, \prd{48}{3235}{93}.}

\bibitem{dv}{S.~Dawson and G.~Valencia, \prd{49} {2188}{94}.}

\bibitem{kl}{D.~Kennedy and B.~Lynn, \np{322}{1}{89}.}

\bibitem{hagi}{K.~Hagiwara {\it et. al.}, \np{282}{253}{87}.}

\bibitem{boud} {F. Boudjema, {\it Proceedings of
Physics and Experiments with Linear
$e^+e^-$  Colliders}, ed. by F.~A.~Harris {\it et al.}, (1993),
p. 713, and references
therein.}

\bibitem{eboli}  {O.~Eboli, M.~Gonzalez-Garcia, and S.~Novaes,
MAD-PH-764, (1993).}

\bibitem{cern}{UA2 Collaboration, J.~Alitti {\it et al.}, \pl{277}{194}{92}.}

\bibitem{lepi}{K.~Hagiwara {\it et al.}, \prd{48}{2182}{93}.}

\bibitem{bauretal}{U.~Baur and E.~L.~Berger, \prd{41}{1476}{90};
U.~Baur, T.~Han, and J.~Ohnemus,  \prd{48}{5140}{93}.}

\bibitem{lepii}{M.~Kuroda, F.~M.~Renard, and D.~Schildknecht,
\pl{183}{366}{87}.}

\bibitem{miya}{For a review, see, {\it e. g.}, A.~Miyamoto, in
{\it Proceedings of Physics and Experiments with Linear
$e^+e^-$  Colliders}, ed. by F.~A.~Harris {\it et al.}, (1993),
p. 141, and references
therein.}

\bibitem{godfrey}{S.~Godfrey and K.~A.~Peterson,
in  {\it Proceedings of Physics and Experiments with Linear
$e^+e^-$  Colliders}, ed. by F.~A.~Harris {\it et al.}, (1993), p. 731.}

\end{thebibliography}
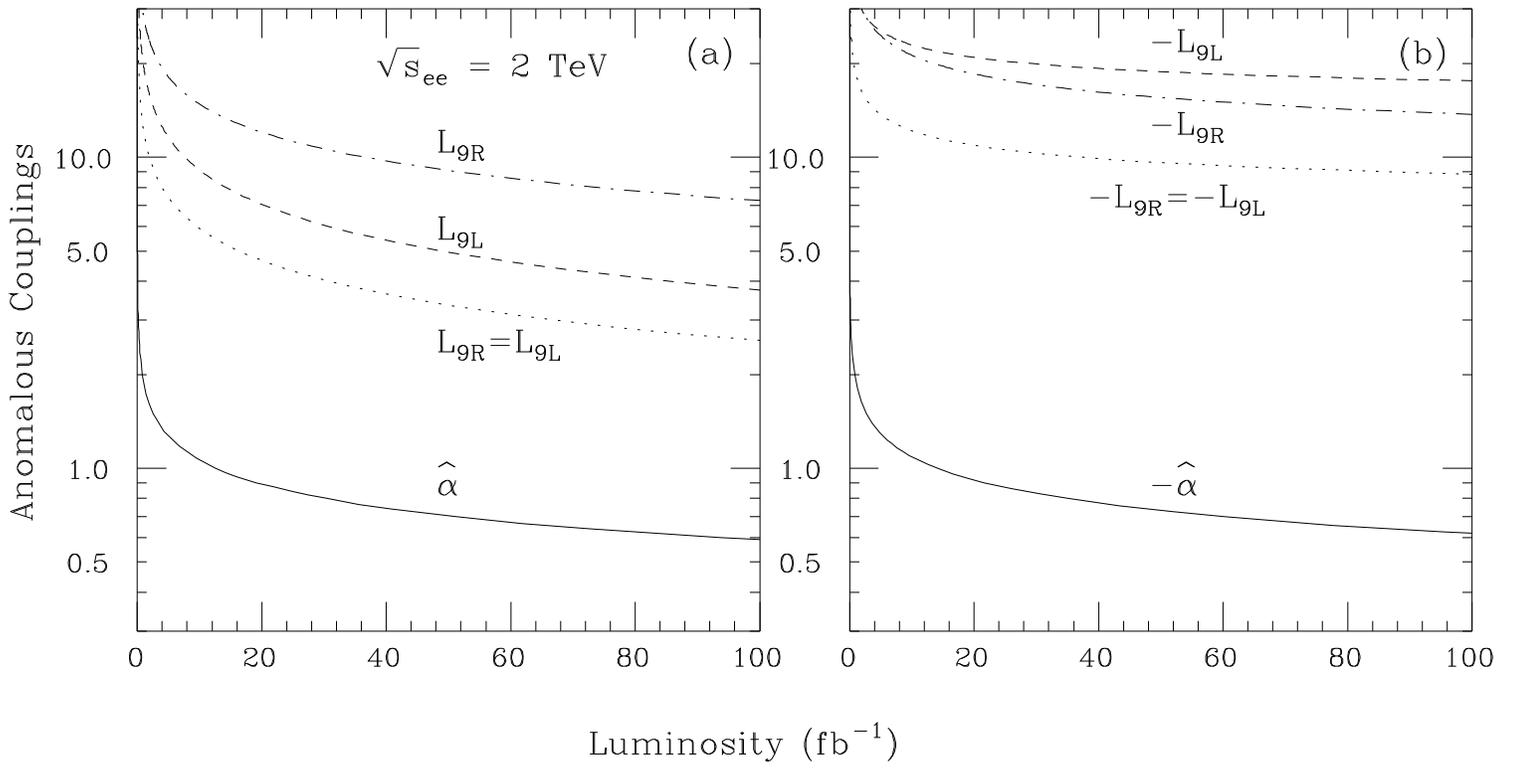
\end{document}